# Isolated Electrostatic Structures Observed Throughout the Cluster Orbit: Relationship to Magnetic Field Strength

J. S. Pickett[1], L.-J. Chen[1], S. W. Kahler[1], O. Santolík[2,1], D. A. Gurnett[1], B. T. Tsurutani[3], and A. Balogh[4]

[1]Department of Physics and Astronomy, The University of Iowa, Iowa City, IA, USA
[2]Faculty of Mathematics and Physics, Charles University, Prague, Czech Republic
[3]Jet Propulsion Laboratory, California Institute of Technology, Pasadena, CA, USA
[4]The Blackett Laboratory, Imperial College, London, UK

Correspondence and Offprint Requests to: J. S. Pickett (pickett@uiowa.edu)

**Abstract**. Isolated electrostatic structures are observed throughout much of the 4 $R_E$ by 19.6 $R_E$ Cluster orbit. These structures are observed in the Wideband plasma wave instrument's waveform data as bipolar pulses (one positive and one negative peak in the electric field amplitude) and tripolar pulses (two positive and one negative peak, or vice versa). These structures are observed at all of the boundary layers, in the solar wind and magnetosheath, and along auroral field lines at 4.5-6.5 $R_E$. Using the Wideband waveform data from the various Cluster spacecraft we have carried out a survey of the amplitudes and time durations of these structures and how these quantities vary with the local magnetic field strength. Such a survey has not been carried out before, and it reveals certain characteristics of solitary structures in a finite magnetic field, a topic still inadequately addressed by theories. We find that there is a broad range of electric field amplitudes at any specific magnetic field strength, and there is a general trend for the electric field amplitudes to increase as the strength of the magnetic field increases over a range of 5 to 500 nT. We provide a possible explanation for this trend that relates to the structures being Bernstein-Greene-Kruskal mode solitary waves. There is no corresponding dependence of the duration of the structures on the magnetic field strength, although a plot of these two quantities reveals the unexpected result that with the exception of the magnetosheath, all of the time durations for all of the other regions are comparable, whereas the magnetosheath time durations clearly are in a different category of much smaller time duration. We speculate that this implies the structures are much smaller in size. The distinctly different pulse durations for the magnetosheath pulses indicate the possibility that the pulses are generated by a mechanism which is different from the mechanism operating in other regions.

## 1 Introduction

There have been numerous reports of observations of isolated electrostatic structures (IES), also referred to as solitary waves and electrostatic solitary waves, and their onsets in the dynamical regions of Earth's magnetosphere (Franz et al., 1998; Kojima et al., 1999; Franz, 2000; Tsurutani et al., 2003; Cattell et al., 2003; Lakhina et al., 2003; and references therein), as well as of theoretical (Schamel, 1986; Muschietti et al., 1999a,b; Chen and Parks, 2002a,b; Chen et al., 2003) and simulation (Omura et al., 1996; Oppenheim et al., 1999; Muschietti et al., 2000; Singh, 2000, Singh et al., 2000) studies of IES. However, no observational studies on how bipolar IES characteristics depend on the local magnetic field strength have been carried out, and there have been no systematic studies on tripolar IES. Most theories of IES provide only analytical solutions for either purely one dimension (Muschietti et al., 1999a,b) or strong magnetic field where the cyclotron radius has been taken as zero (Schamel, 1986; Muschietti et al., 2002; Chen and Parks, 2002a,b; Chen et al., 2003).  On the other hand, in three dimensional unmagnetized plasmas, it has been shown that fully localized solitary wave solutions do not exist if the ambient plasma is isotropic (Chen, 2002).  Several simulation studies have shown that in more than one dimension, bipolar IES are short lived if the magnetic field is zero or so weak that the cyclotron frequency is less than the plasma frequency (Morse and Nielson, 1969; Oppenheim et al., 1999; Muschietti et al., 2000).  In space, however, bipolar IES have been observed in much weaker magnetic fields than what is thought possible by simulation and theoretical studies.  In this paper, using data from the Cluster spacecraft, we report on a statistical survey of the dependence of electric field amplitudes and time durations of bipolar and tripolar IES on the magnetic field strength.

Cluster began its mission operations phase on 1 February 2001 following a five-month commissioning phase.  The Cluster quartet is in an approximate polar orbit with an apogee of about 19.6 $R_E$ and a perigee of 4 $R_E$.  This allows Cluster to cross all of Earth's boundaries and spend considerable amounts of time in the near-Earth solar wind, magnetosheath, magnetotail and auroral zone.  The Cluster orbit precesses through all 24 hours of Magnetic Local Time (MLT) in one Earth-year.  Thus, the Cluster spacecraft are always in the same local time sector for any given month each year as described in Escoubet et al. (2001) and as depicted in their Figure 1.  Although not important for the current study, the inter-spacecraft separations of the four satellites have been changed on a 6- to 12-month time frame since operations began, with the smallest ideal tetrahedron separation being 100 km and the largest 5,000 km as of this date.

The Cluster WBD instrument obtains high resolution waveforms in three different filter bandwidths, providing time resolution between samples varying from 5 to 36.5 µsec (see Gurnett et al., 1997 and Pickett et al., 2003 for a full description of the WBD instrument and its sampling characteristics). This allows for the resolution of periodic or single period (pulse) waveforms with extremely short durations. Since WBD makes only a single axis electric field measurement, it is not possible to transform the waveforms into any other coordinate system. However, it is possible to transform other Cluster three axis measurement and position vectors into the antenna coordinate system used by WBD, thus providing the antenna's position at any moment in time in some reference system, such as a magnetic field-aligned coordinate system. The WBD instrument is unique among all the Cluster instruments in that it does not record its data onboard. Rather, the spacecraft transmits the WBD data in real time to a Deep Space Network ground station. This allows WBD to obtain extremely high time resolution measurements with tens of µsec accuracy in absolute time. In order to obtain these unique capabilities, WBD data are only obtained for a total of 8 hours across all 4 spacecraft (or about 2 hours per spacecraft) for each 57-hour Cluster orbit.

The combination of the Cluster orbit characteristics and the WBD instrument on each spacecraft provides the unique capability for the first time to obtain high resolution digital waveform data from multiple spacecraft from all regions near Earth in which isolated electrostatic structures are observed. Although wave instruments on several single spacecraft such as Freja, Geotail, WIND, Polar and FAST have observed IES, none of them explore as many of the regions where IES are observed as Cluster does.

Furthermore, most of the waveform instruments on these spacecraft are unable to observe the extremely short duration IES as WBD does, or those that do cannot sample as many of the IES regions.  As will be shown below, IES are observed along magnetic field lines that map to the auroral zone, in the magnetosheath and near-Earth solar wind, and at all boundary layers.  In general, wherever there is turbulence or a mixing of plasmas, IES are observed.

Below we begin by providing samples of the IES of interest.  We follow this by the results of the survey, specifically the amplitude and time duration of the IES vs. the magnetic field strength, and end with some possible explanations for the observed features.

## 2  IES Examples

Figure 1a shows a typical 6-minute spectrogram of data obtained by WBD on 12 June 2002 on all four spacecraft in the auroral zone at about 4.8 $R_E$ and 53° magnetic latitude. This spectrogram has increasing time, in UT, plotted on the horizontal axis and frequency, in kHz, on the vertical axis with color indicating power spectral density, in $V^2/m^2/Hz$.  The spectrogram was created by taking 1024 samples of the time series and transforming these data to the frequency domain by using a Fast Fourier Transform.  The white line in each panel is the electron cyclotron frequency as determined from the Fluxgate Magnetometer, FGM (see Balogh et al., 1997).  Figure 1b shows a 25 msec line plot of the waveform beginning at 02:48:42.546 UT obtained by WBD on C4 (Cluster 4) from the 6-minute interval seen in the spectrogram. The line plot in Figure 1b has increasing time, in sec from 02:48:42.546 UT, plotted on the horizontal axis and electric field amplitude, in mV/m, plotted on the vertical axis. The total angle of the electric field antenna to the local magnetic field using transformed FGM data is shown on the right vertical scale. The spectrogram at the time of the waveform in Figure 1b shows only a broad band signal ranging in frequency from the lower cutoff of the filter around 50 Hz, where its greatest intensity is observed, up to its higher cutoff around 19 kHz, where a much lower intensity is observed. The broad band signal results from the fact that the pulses observed in the waveform in Figure 1b contain all frequencies. When one or more of these pulses are dominant in a 1024 point sample and are transformed to the frequency domain via Fast Fourier Transform, the expected result is a broad band signal as observed.

Two types of IES are pointed out in Figure 1b. The first is a tripolar pulse, defined as having either two positive peaks and one negative peak in the electric field, as in this case, or two negative peaks and one positive. This tripolar pulse has a fairly typical duration of about 2 msec and amplitude of 55 mV/m peak-to-peak. As shown in Pickett et al. (2004), when the electric field of these pulses is integrated over the duration of the pulse, a significant potential shift is obtained. The second type of pulse pointed out in Figure 1b is the bipolar pulse (one positive and one negative peak). Likewise, the bipolar pulses have typical durations of 760 μsec and amplitudes of 75 mV/m peak-to-peak. No potential shift will result from integrating the electric field of the bipolar pulse over the duration of the pulse. Figure 1b also shows that the total angle of the electric antenna being used by WBD to the local magnetic field is around 157º, or nearly antiparallel.

We now highlight some of the differences between the IES in the near-Earth auroral zone presented in Figure 1, and the magnetosheath. Figure 2 (same format as Figure 1 with the exception that the electron cyclotron frequency, plotted as a white line, has been omitted in Figure 2a as it would appear at the very bottom of the frequency band) contains WBD data taken during a pass through the dayside magnetosheath at 13-14 $R_E$ and 20º-30º magnetic latitude. Figure 2a shows the spectrogram for a two-hour period of time on 6 April 2002 when the spacecraft were separated by about 100 km. Note that the horizontal lines seen around 40 kHz and 65 kHz in all panels are spacecraft interference and the numerous horizontal lines seen in the C1 and C3 panels are interference from the Cluster EDI experiment. Although the signatures of the IES, i.e., the broad band signals, in this spectrogram, are observed to be very similar from one spacecraft to the next, we have been unable to conclusively correlate any one structure as propagating from one spacecraft to the next, but this is the subject of ongoing work and not pertinent to this study. Here we see that the broad band signals extend up to about 40-60 kHz, which is indicative of the pulses being much shorter in duration (see Pickett et al., 2003) than those from the auroral zone. Indeed, when we look at the 5 msec sample waveform from C4 beginning at 22:25:01.2566 UT (Figure 2b), we see that the duration of these bipolar pulses is 54 μsec. The interesting aspect of these bipolar pulses is that they are seen almost every 1.4 msec at a certain phase of the lower frequency, nearly sinusoidal, wave. This would imply that the pulses are being created through some mechanism that involves the low frequency wave, but this topic is also left to future work. In addition we observe that the amplitudes of the pulses, 0.02 to 0.10 mV/m, are much lower than in the auroral zone.

## 3 IES Survey

Since the primary purpose of this paper is to provide a geophysical basis, in this case the local magnetic field strength, for the various amplitudes and time durations of the pulses, we resort to a survey of the IES observed in various parts of the Cluster orbit. Because it would be computationally prohibitive at this time to run an automatic IES detection program on all of the data obtained to date, we have instead turned to choosing representative events from the various regions that Cluster traverses in which IES are observed, and only then analyzing the first second of every minute of data.

The automatic bipolar and tripolar IES detection algorithm works as follows. First the range of the background noise is determined by taking the standard deviation of the raw count values over 2 or 3 minor frames of data (approximately 80-120 μsec). Then looking within this span of data, the program finds the first data point that falls outside the noise range. For this data point, a local minimum and maximum raw count value are found. The following tests are now run to determine if the data represent a bipolar pulse:

1. Ensure that the maximum and minimum raw count values for the pulse do not occur at the beginning or ending of the pulse,

2. Ensure that the slope of a linear best-fit line between the peaks of the pulse is greater in magnitude than the slope of a linear best-fit line between the start and the first peak and the slope for the line between the second peak and the end of the pulse.

3. The ratio of the amplitudes for the two peaks (amplitude measured from the mean of the pulse) must be smaller than 2:1.

4. There are no significant raw count values within 1/4 the period of the pulse. A significant value is one who's amplitude (measured from the mean of the pulse) is greater than 10% of the pulse's peak-to-peak amplitude.

5. The time duration of each peak must be roughly the same. The largest ratio accepted for the time duration of each individual peak is 14:11.

If all the above five conditions are met, the pulse is confirmed to be a bipolar pulse. If any one of conditions 3 through 5 is NOT met, then another local minimum or maximum (i.e., if the second peak found previously was a minimum, then we are looking for a maximum) is identified. The following tests are now run to determine if the data represent a tripolar pulse:

1. Ensure that the maximum and minimum raw count values for the pulse do not occur at the beginning or ending of the pulse.

2. Ensure that the slopes of a linear best-fit line between the first and second peaks and the line between the second and third peaks are the greatest in magnitude for the pulse, i.e., the same as test applied above for bipolar pulses.

3. Ensure that the second peak (center peak) is the farthest removed from the mean raw count value of the pulse.

4. There are no significant raw count values within 1/4 the period of the pulse. A significant value is one who's amplitude (measured from the mean of the pulse) is greater than 10% of the pulse's peak-to-peak amplitude.

5. The amplitudes (measured from the mean of the pulse) of the two peaks on one side of the mean of the pulse individually should at least be 1/4 the amplitude of the single pulse on the other side of the mean.

6. The time duration of the two peaks on the same side of the pulse's mean raw count value must be similar. The largest ratio accepted for the time duration comparison of these two individual peaks is 3:1.

If all the above six conditions are met, then the pulse is confirmed to be a tripolar pulse. This method of pulse detection will not result in false positives. However, because of the various fluctuations naturally present in the data, i.e., other plasma waves and electrostatic turbulence, some pulses may be missed. Since many of these missed pulses are questionable, they are best left out of this survey. The pulses that are confirmed have been extensively checked manually for authenticity. In addition, the confirmed pulses are also labeled as "clipped" if the instantaneous dynamic range of the instrument is insufficient to provide an accurate amplitude. In the plots that follow below, all "clipped" pulses have been omitted.

The results of the IES survey are shown in Figures 3 and 4 for the bipolar and tripolar pulses, respectively. These plots contain 7,104 confirmed bipolar pulses and 668 confirmed tripolar pulses. First we examine the bipolar pulses. Figure 3a shows a scatter plot of the amplitude, in mV/m, on the horizontal axis vs. the total magnetic field strength, in nT, on the vertical axis, color coded according to the region in which the bipolar pulses were detected. Note that the region labeled Plasma Sheet is applicable only for distances of 18-19 $R_E$ and that the Polar Cap represents crossings of the polar cap boundary layer at two different distances, 5 and 10 $R_E$. For all regions we have plotted an "x" within a bracketed line. These are the mean values and standard deviations, respectively, of the points within that region. Although there is a range in the observed amplitudes of the bipolar pulses at any given magnetic field strength, the trend for the amplitude to be larger as the strength of the magnetic field increases is clearly evident through the provided values of mean and standard deviation. We will present our attempt to understand this trend in the next section. From this figure we also see that the range of amplitudes varies from a few hundredths mV/m to several tens of mV/m. Since WBD employs an automatic gain function implemented in its hardware to add and remove gain in 5 dB steps at an update rate of once every tenth of a second, it should not be missing many structures over the 1 second interval because the gain will adjust 9 times in this period if necessary to keep the amplitudes within its dynamic range. If, however, all of the gain has been taken out of the system and the structures are still being clipped, WBD would be unable to determine the full amplitude. As can be seen from Figure 3a, this is most likely to occur in the auroral zone where the largest magnetic fields are observed for the Cluster orbit.  As stated previously, structures with clipped waveforms are not plotted in Figure 3.

The survey of the time durations of the same sample of bipolar pulses is shown in Figure 3b.  Here we have plotted the time duration of the pulses on the horizontal scale, in msec, vs. the strength of the magnetic field, in nT, on the vertical axis, color coded by region with mean and standard deviation values provided within each region.  We observe that there is no dependence of the time duration of the pulses on the magnetic field.  However, we do observe that there appears to be two distinct classes of time durations:  1) those greater than 100 μsec up to about 5 msec, and 2) those less than 100 μsec down to the lowest duration that WBD can resolve at about 20 μsec.  Clearly the magnetosheath structures fall into category 2 and all others into category 1.  The reason for this distinction will be discussed in the next section and simply point out here that for a typical flow velocity of a few hundred km/sec in the magnetosheath and a time duration less than 100 μsec, we obtain a scale size on the order of tens of meters.

We have also examined the tripolar pulse amplitudes and time durations for dependences on the magnetic field strength.  The survey results are provided in Figures 4a and 4b (same format as for Figures 3a and 3b).  Although fewer tripolar structures are detected in any sample interval, similar trends, or lack thereof, to the bipolar structures are seen, i.e., the tripolar pulse amplitudes increase as the magnetic field strength increases over a finite spread of amplitudes, and the time durations show no dependence on the magnetic field strength.  However, just as for the bipolar pulses, the tripolar pulses in the magnetosheath are from a different, shorter time duration class.  Similar results for the tripolar pulses as for the bipolar ones point to the conclusion that they are coherent structures of the same nature but of different shapes.  The fact that the tripolar pulses are sometimes observed in groups without the presence of bipolar pulses (see, for example, Figure 2b of Pickett et al., 2004) indicates that they are themselves coherent structures and less likely to be an intermediate stage in the generation of bipolar pulses.  There is no apparent reason for why there are more bipolar than tripolar pulses (10:1) detected in any given sample interval.  However, it is the case that the automatic detection algorithm will reject more possible cases of tripolar than bipolar pulses due to the added complexity of cleanly detecting a third peak.

## 4  Discussion

In this section, we discuss our attempts to understand the observational features shown in Figures 3(a)-4(b).  The large spread in the electric field amplitudes for a fixed magnetic field strength is consistent with the key property of Bernstein-Greene-Kruskal (BGK) solitary waves whose widths and potential amplitudes are constrained by inequalities (Chen and Parks, 2002a,b; Muschietti et al., 2002; Chen et al. 2003).  For a given size of a BGK solitary structure, there is a continuous range of allowed potential amplitudes, and vice versa.  Therefore, a large range of continuous electric field amplitudes (potential amplitudes divided by size) are naturally expected for BGK solitary waves.  On the other hand, for fluid solitons such as KdV solitons (Washimi and Taniuti, 1966), their widths increase with decreasing amplitudes.  For a certain amplitude, there is only one allowed width for a fluid soliton, and as such, it is less likely that there is a wide range of electric field amplitudes for a particular geomagnetic region.

There are so far no theories that describe how tripolar pulses are self-consistently supported by a collisionless plasma, how they are generated, and what their properties are, and no knowledge about how they would behave in finite magnetic fields. For the tripolar pulses themselves to be coherent structures, trapping of both ions and electrons must be involved (Pickett et al., 2003). We therefore extrapolate that their size and amplitudes are also constrained by inequalities based on the BGK theory of Chen and Parks (2002a,b) and Chen et al. (2003) for bipolar pulses that require either ion trapping or electron trapping. The large span in electric field amplitudes is consistent with our extrapolation.

How the properties of solitary waves vary with the strength of the magnetic field is still poorly addressed by theories. The width-amplitude inequality relations for BGK solitary waves derived in the strong magnetic field limit have been shown to hold at least in the regime where the cyclotron radius of the particles trapped in the solitary structure is much less than the size of the structure (Chen et al., 2003). This condition is based on the stability requirements of a BGK solitary wave when the lowest order effects of a finite magnetic field are considered, and can be written as the following:

$$\frac{\sqrt{m\Psi/e}}{B\delta_z} << 1 \qquad\qquad (1)$$

$$\frac{\sqrt{2m\Psi/e}}{B\delta_r} << 1 \qquad\qquad (2)$$

where $\Psi$ is the peak potential amplitude, B is the magnetic field strength, $\delta_z$ is the parallel size, $\delta_r$ is the perpendicular size, and e,m are the charge and mass of particles (electrons for electron mode solitary waves, and ions for ion mode), respectively. The above conditions point to a trend that for a much weaker magnetic field, either the potential amplitude would decrease or the size would increase in order for the structures to be stable, and this results in smaller electric field amplitudes. The Cluster observational results shown in Figures 3(a) and 4(a) are consistent with this, i.e., the typical electric field amplitude for a particular magnetic field strength tends to decrease with decreasing magnetic fields.

We now discuss the possible underlying reason for the feature that the magnetosheath pulse durations form a different class while the rest of the pulses forms a group with no dependence on the magnetic field strength. The time duration that a pulse is observed scales with the ratio of its size to its velocity. If the pulses are generated by the two-stream instability, this ratio would be roughly constant with a certain spread. The reason is as follows. The typical size of these pulses is determined by the wavelength of the fastest growing mode in the linearly unstable stage, and this wavelength is proportional to the relative streaming velocity of the two beams (Gurnett and Bhattacharjee, 2003). On the other hand, the velocities of the pulses generated by the two-stream instability are roughly half of the two stream velocity with some deviation, if the observing frame is taken to be one of the two streams. Hence, the average ratio of size to the velocity of the pulses would stay roughly the same for different regions if they are generated by unstable two streams even with different relative velocities. In this scenario, the fact that the magnetosheath pulse durations themselves form a different category would then indicate that these pulses are not generated by a two-stream instability.

What mechanism could generate the magnetosheath pulses? As suggested by Chen et al. (2003), BGK solitary waves may be spontaneously generated by turbulence in the absence of a two-stream instability. We suspect that the magnetosheath pulses are examples of such spontaneous generation since the magnetosheath is known to be rich in turbulence. If this is the case, the much shorter time durations of magnetosheath pulses may just reflect the scale lengths of the turbulence there. We note that although the magnetosheath pulses may well be generated by a different mechanism, once they are generated, the conditions for their stability are the same as the pulses in other regions. Therefore, the magnetosheath pulses are not expected to stand out from the trend in the electric field amplitude vs. magnetic field plots.

**5 Conclusions**

We have presented new observational features of isolated electrostatic structures in finite magnetic fields that call for new theories. We have observed electric field structures in the form of bipolar and tripolar pulses in all of the following regions crossed by the Cluster quartet: numerous boundary layers, the near-Earth solar wind, magnetosheath, magnetotail and auroral zone.  These IES have amplitudes ranging from a few hundredths of mV/m up to almost 100 mV/m, and time durations from about 20 μsec, the shortest time duration that can be resolved, up to about 5 msec.  The electric field amplitudes of the IES show a wide range of scattering even within a given geomagnetic region, and exhibit a trend of increasing electric field amplitudes with increasing strength of the magnetic field.  No dependence is found for the duration of the IES pulses on the strength of the magnetic field.  However, plotting the time durations of the IES vs. the magnetic field strength clearly points out that the magnetosheath IES come from a different class of structures than those found in all other regions.  We have speculated that the IES are BGK mode and that the trend in the electric field/magnetic field comes from the stability requirements of the BGK mode in finite magnetic fields.  The lack of dependence of the time duration (or size/velocity) is indicative of the generation mechanism in most regions (except the magnetosheath) being the two-stream instability.  The difference in the time durations of the magnetosheath IES vs. all other observed IES is most likely due to a much different generation mechanism.

*Acknowledgments*.  This work was supported under NASA/Goddard Space Flight Center Grant No. NAG5-9974 and NSF Grant ATM 03-27450.  We thank all of the many groups on the European side for their part in obtaining the WBD data (ESA, ESTEC, ESOC, JSOC, Sheffield University, and the Cluster Wave Experiment Consortium), as well as those on the U.S. side (JPL/DSN).  We also thank ESA and NASA for valuable analysis support tools, namely, CSDSWeb, SSCWeb and CDAWeb).  Portions of this work were performed at the Jet Propulsion Laboratory, California Institute of Technology under NASA contract.

**Figure Captions**

Figure 1: Cluster WBD data taken on 12 June 2002 in the auroral zone.  a)  Spectrogram showing the frequency and power spectral density of the emissions.  The broad band signals ranging up to about 10 kHz are indicative of times when IES are observed.  b) Representative waveform from a time of the broad band signals, showing the two types of IES:  bipolar and tripolar pulses.

Figure 2:  Cluster WBD data taken on 6 April 2002 in the dayside magnetosheath.  a) Spectrogram showing the frequency and power spectral density of the emissions.  The broad band signals in this region, unlike the auroral zone, range in frequency from the lowest frequency measured up to 40-60 kHz.  b) Representative bipolar pulse waveforms from a time of the broad band signals, showing that the pulses are much shorter in duration in this region and that they are tied to a certain phase of the low frequency wave.

Figure 3:  Survey of the bipolar pulses observed by Cluster WBD over a two-year period. a) Electric field amplitude vs. magnetic field strength showing a trend of increasing electric field amplitude with increasing magnetic field strength.  The over plotted bracketed lines with an imbedded "x" within each regional grouping represent the standard deviation and mean of that group, respectively.  b)  Pulse duration vs. magnetic field strength showing no trend between the two, but pointing out the obvious difference of the time durations of the magnetosheath pulses to pulses in all other regions.

Bracketed lines with the imbedded "x" represent the standard deviation and mean of each regional group.

Figure 4:  Survey of the tripolar pulses observed by Cluster WBD over a two-year period. a) Electric field amplitude vs. magnetic field strength showing a trend of increasing electric field amplitude with increasing magnetic field strength, just as for the bipolar pulses. b) Pulse duration vs. magnetic field strength showing no trend between the two, but pointing out the obvious difference of the time durations of the magnetosheath pulses to pulses in all other regions, just as for the bipolar pulses.

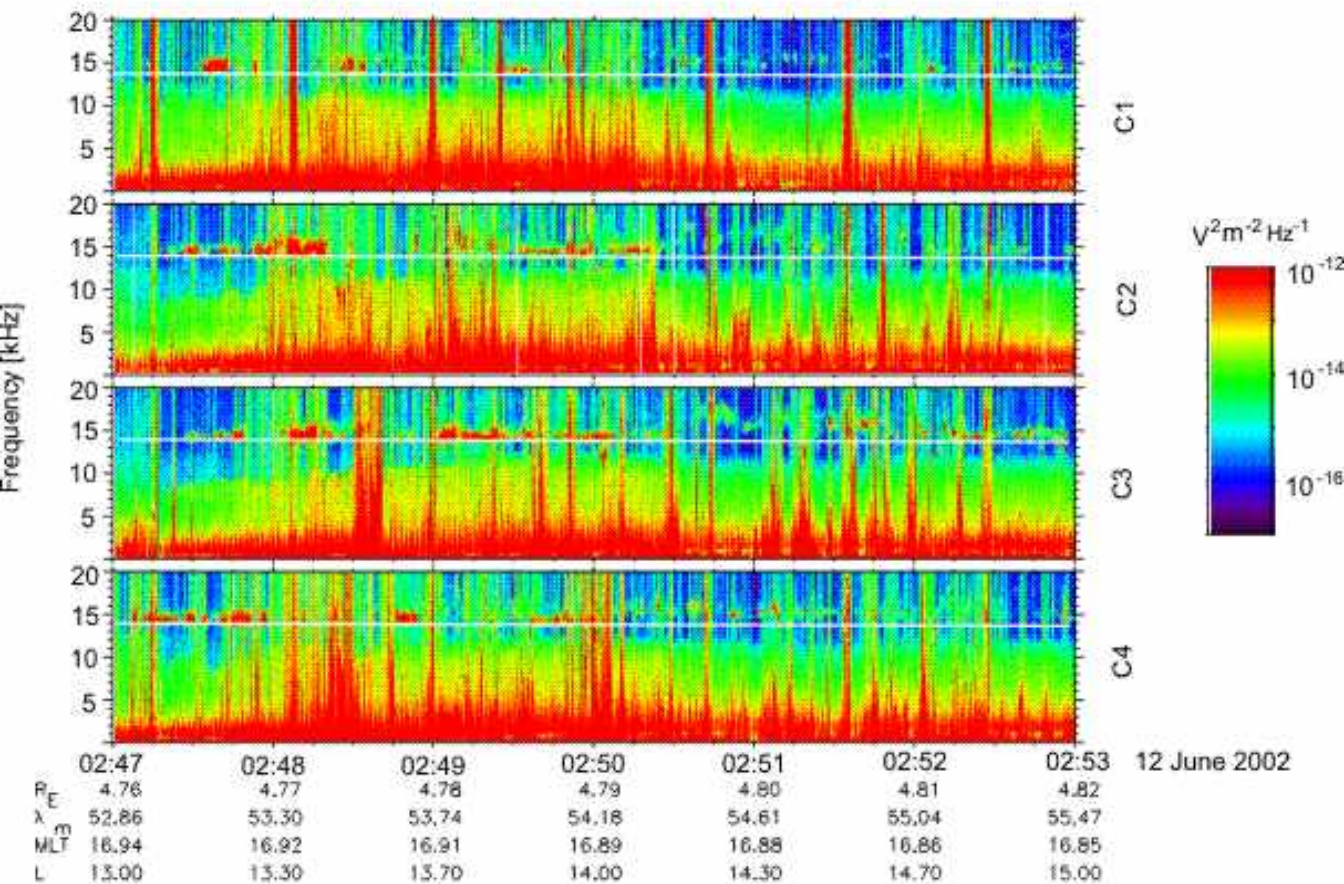

Figure 1a

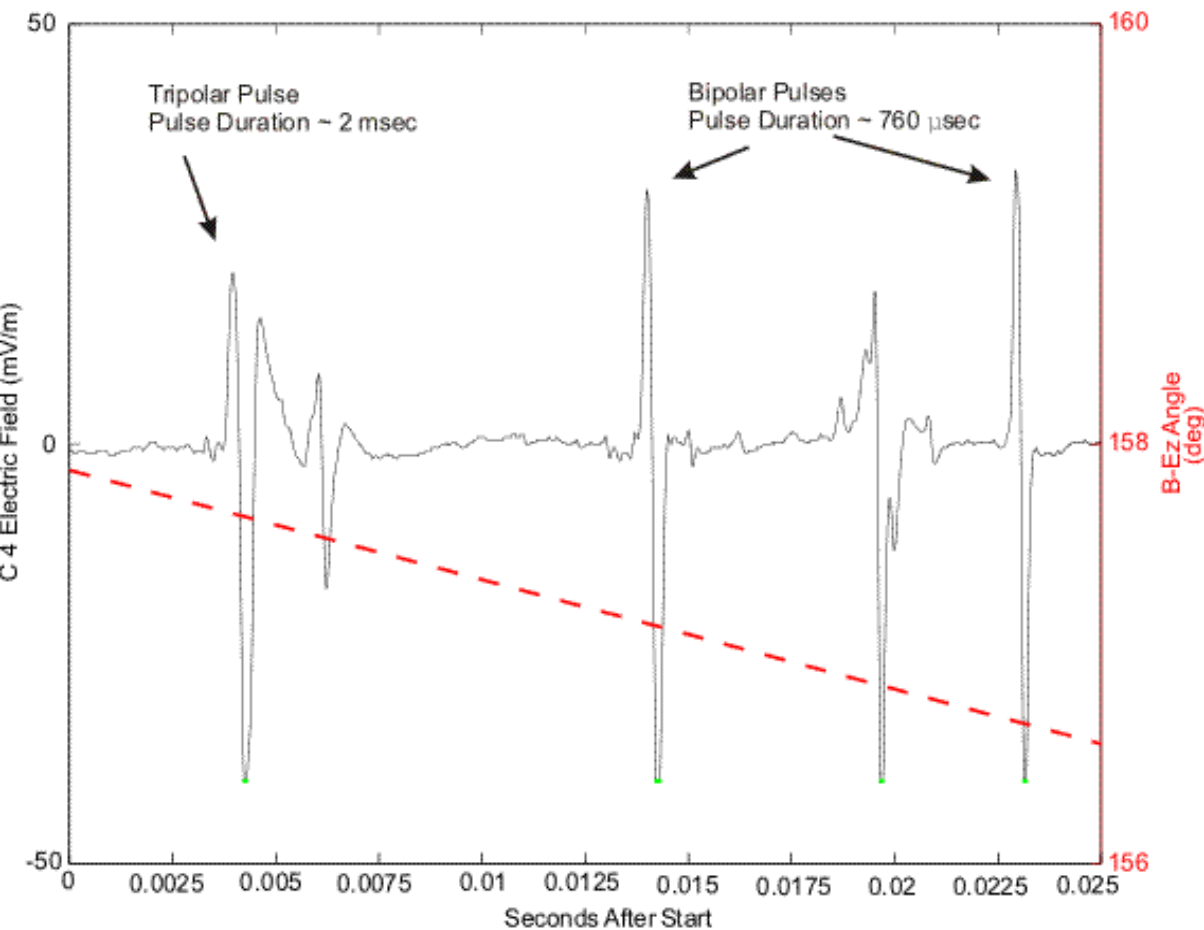

AURORAL ZONE (4.8 $R_e$ 53.5 $\lambda_{inv}$, 16.9 MLT, 13.5 L)

Tripolar Pulse
Pulse Duration ~ 2 msec

Bipolar Pulses
Pulse Duration ~ 760 μsec

C 4 Electric Field (mV/m)

B-Ez Angle (deg)

Seconds After Start

START: 02:48:42.546 on 12 June 2002
Figure 1b

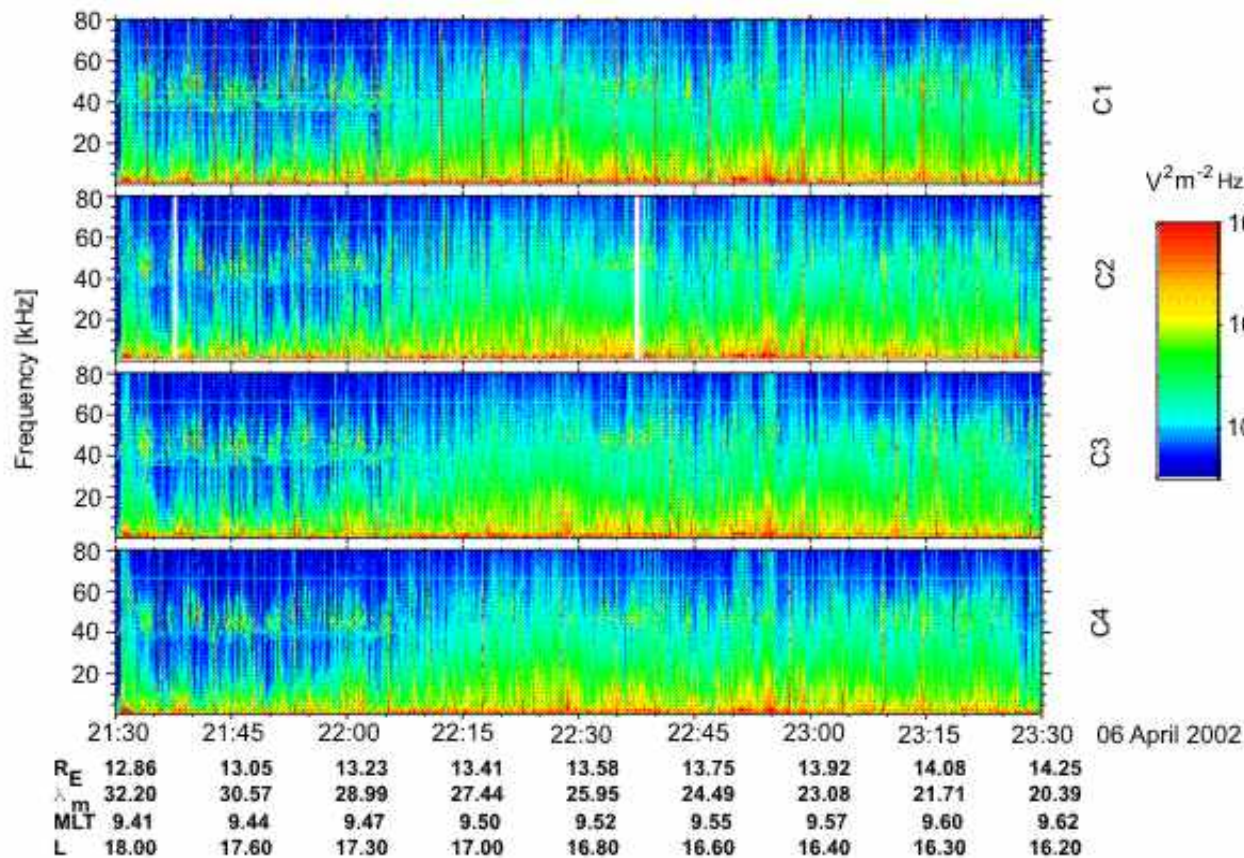

**CLUSTER WBD**
**DAYSIDE MAGNETOSHEATH**
(SPACECRAFT SEPARATION ~ 100 KM)

Figure 2a

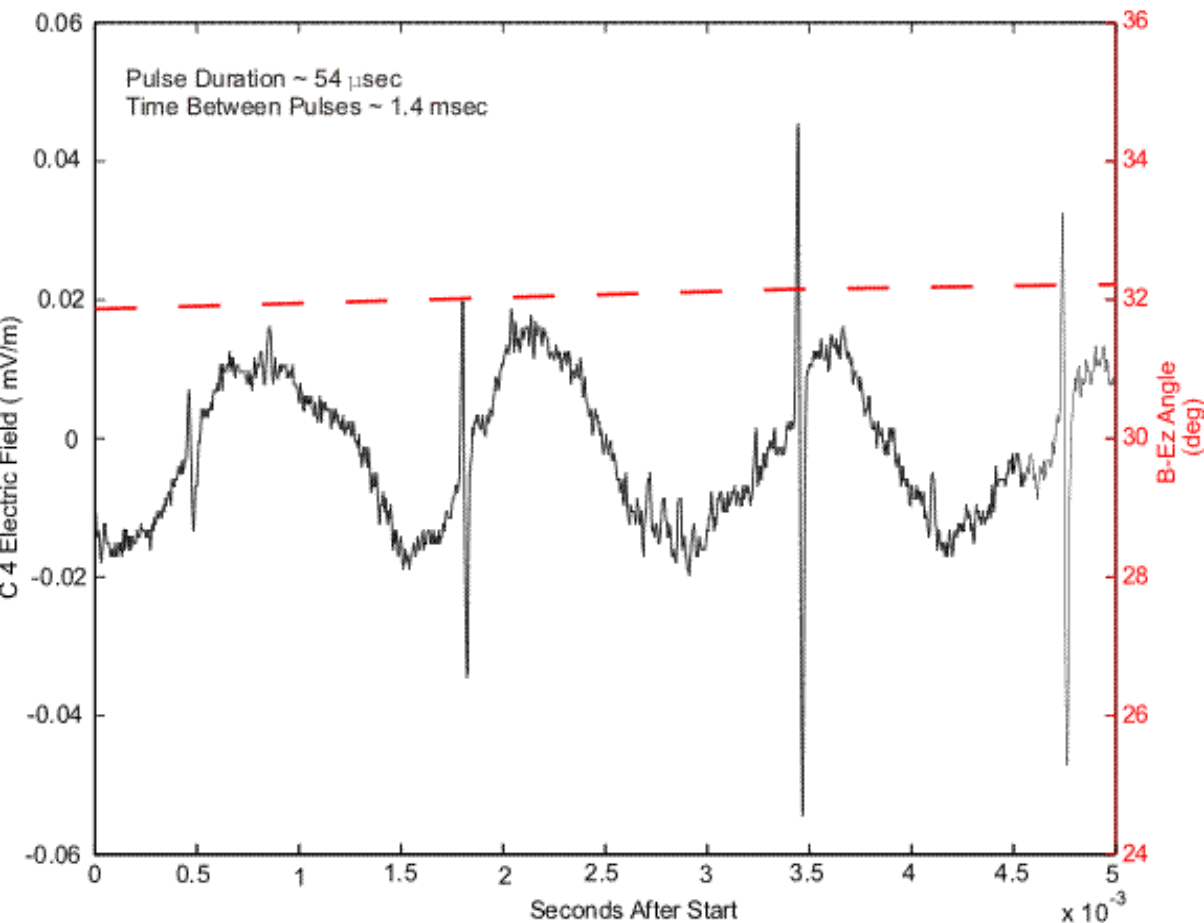

MAGNETOSHEATH (13.4 $R_e$, 26.4 $\lambda$, 9.5 MLT)   BIPOLAR PULSES

Pulse Duration ~ 54 µsec
Time Between Pulses ~ 1.4 msec

START: 22:25:01.2566 UT on 06 April 2002

Figure 2b

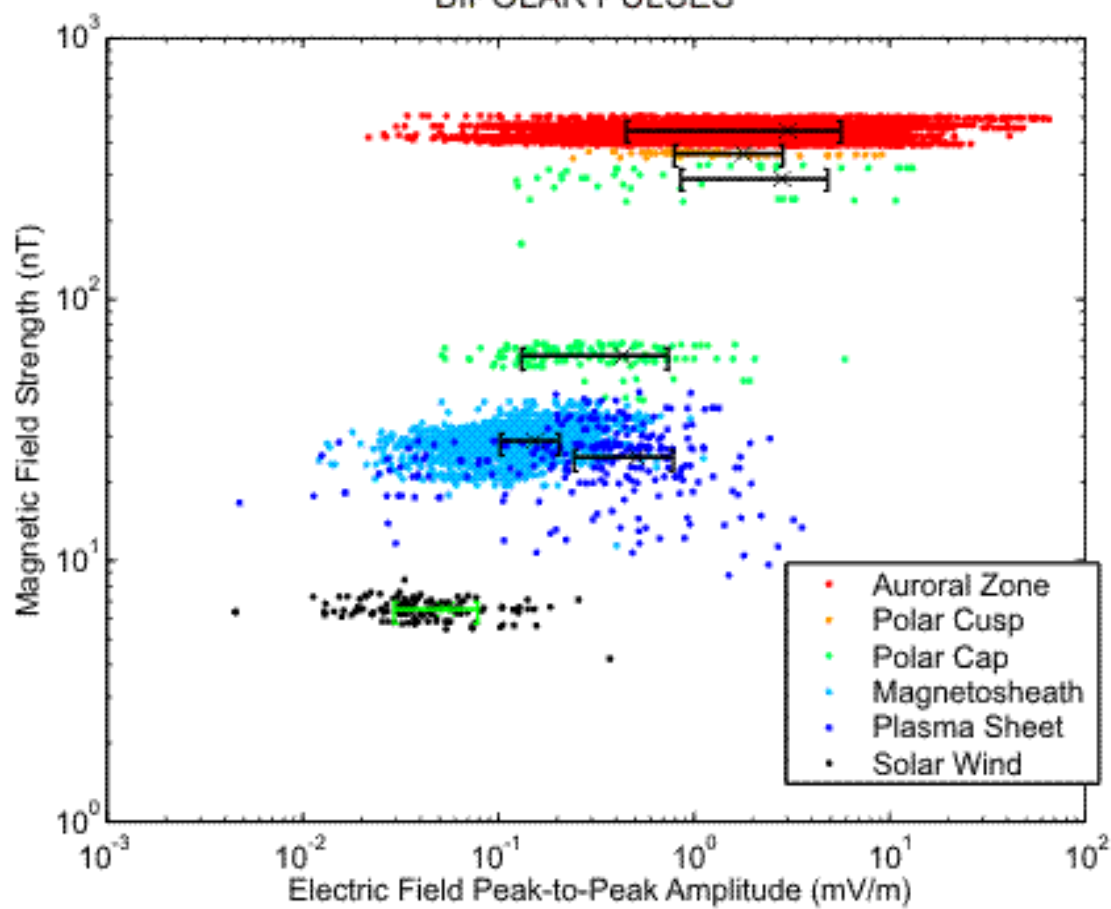

Figure 3a

BIPOLAR PULSES

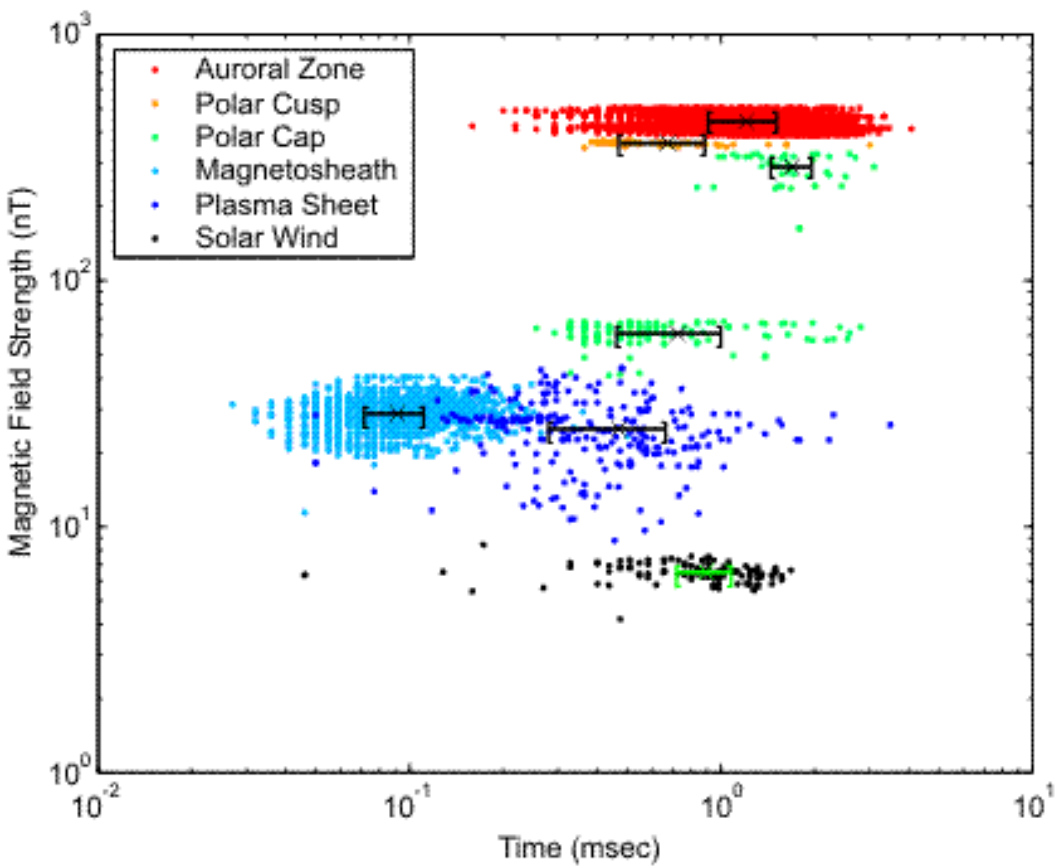

Figure 3b

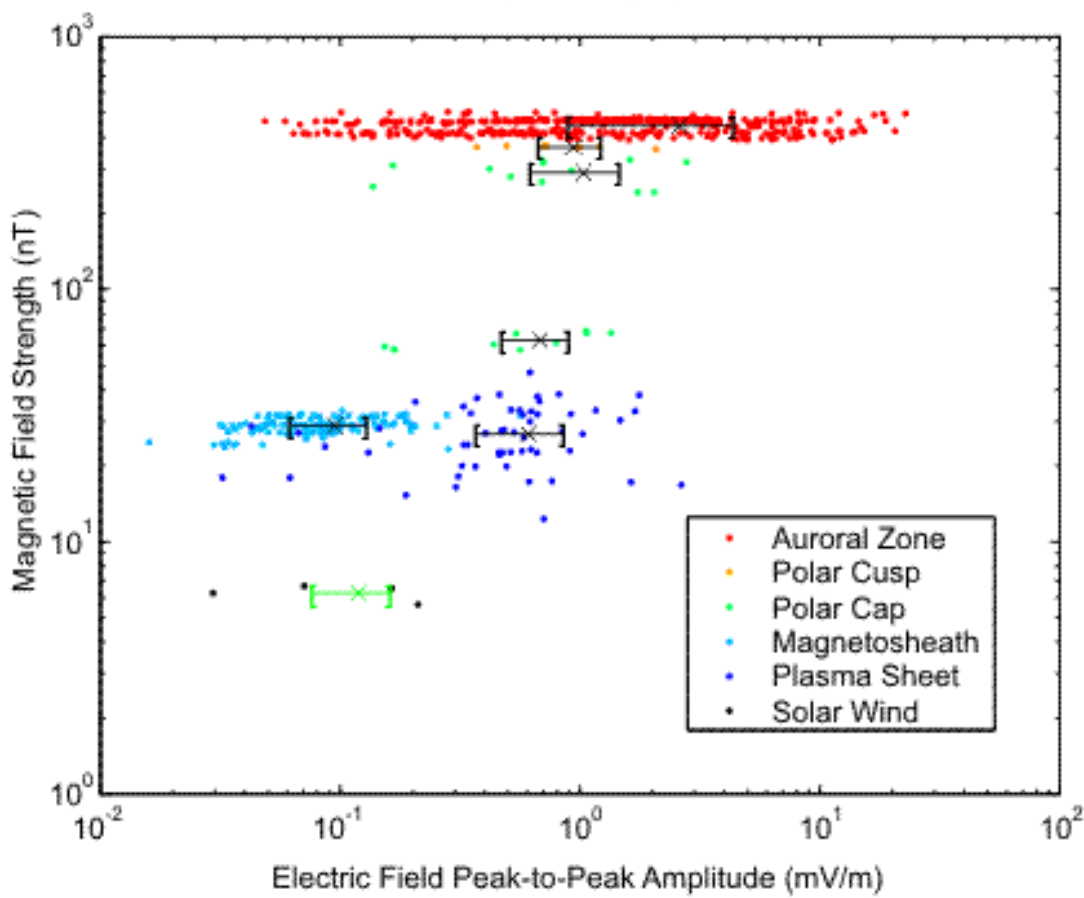

Figure 4a

TRIPOLAR PULSES

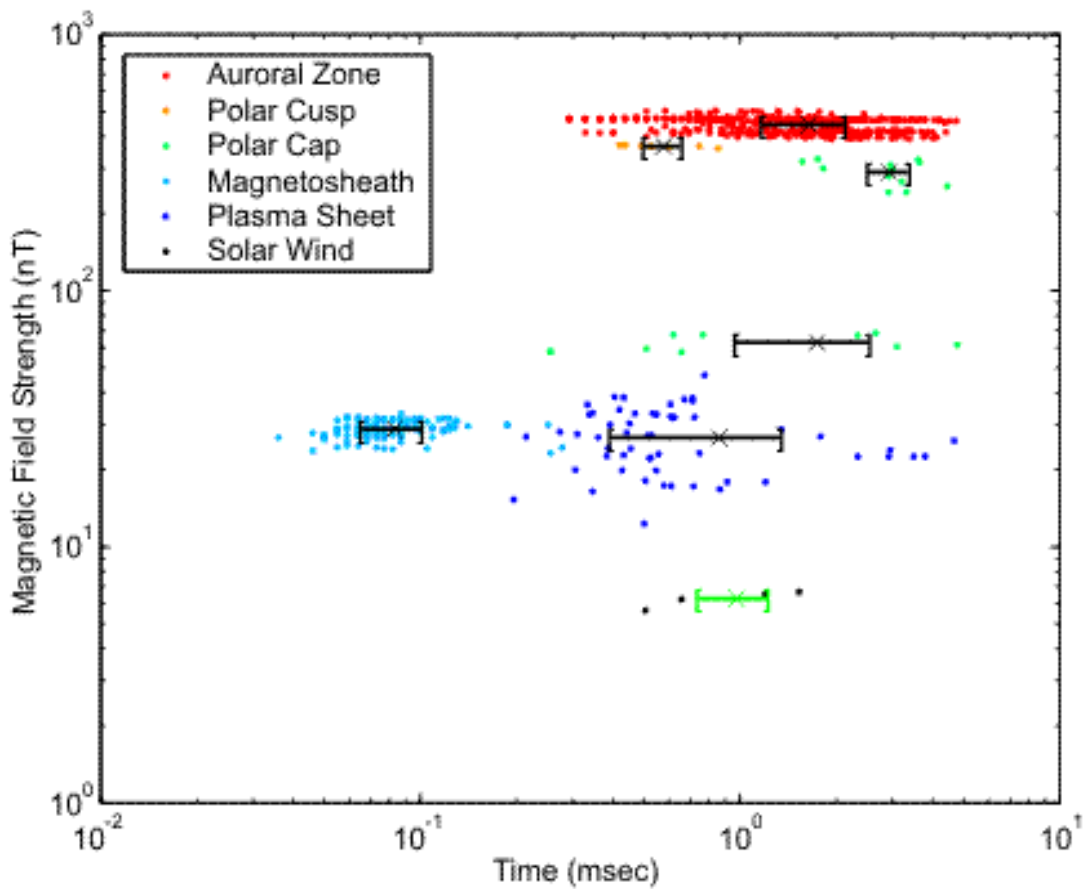

Figure 4b